\documentclass[aps,pra,showpacs,twocolumn,superscriptaddress]{revtex4-1}
\usepackage{epsfig}
\usepackage{graphicx}
\usepackage{dcolumn}
\usepackage{color}

\begin{document}

\title{Radio-Frequency-to-Optical Conversion using \\Acoustic and Optical Whispering Gallery Modes}

\author{Rekishu Yamazaki}
\email[]{rekishu@icu.ac.jp}
\altaffiliation[Current address: ]{College of Natural Sciences, International Christian University, Mitaka, Tokyo 180-8585, Japan}
\affiliation{Research Center for Advanced Science and Technology~(RCAST), The University of Tokyo, 4-6-1 Komaba, Meguro-ku,
	Tokyo 153-8904, Japan}
\affiliation{PRESTO, Japan Science and Technology Agency, Kawaguchi-shi, Saitama 332-0012, Japan}

\author{Ayato Okada}
\affiliation{Research Center for Advanced Science and Technology~(RCAST), The University of Tokyo, 4-6-1 Komaba, Meguro-ku,
	Tokyo 153-8904, Japan}

\author{Atsushi Noguchi}
\affiliation{Research Center for Advanced Science and Technology~(RCAST), The University of Tokyo, 4-6-1 Komaba, Meguro-ku,
	Tokyo 153-8904, Japan}
\affiliation{PRESTO, Japan Science and Technology Agency, Kawaguchi-shi, Saitama 332-0012, Japan}

\author{Shingo Akao}
\affiliation{Ball Wave Inc.,  Sendai, Miyagi 9890-8579, Japan}

\author{Yusuke Tsukahara}
\affiliation{Ball Wave Inc., Sendai, Miyagi 9890-8579, Japan}

\author{Kazushi Yamanaka}
\affiliation{Ball Wave Inc., Sendai, Miyagi 9890-8579, Japan}

\author{Nobuo Takeda}
\affiliation{Ball Wave Inc., Sendai, Miyagi 9890-8579, Japan}

\author{Yutaka Tabuchi}
\affiliation{Research Center for Advanced Science and Technology~(RCAST), The University of Tokyo, 4-6-1 Komaba, Meguro-ku,
	Tokyo 153-8904, Japan}

\author{Koji Usami}
\affiliation{Research Center for Advanced Science and Technology~(RCAST), The University of Tokyo, 4-6-1 Komaba, Meguro-ku,
	Tokyo 153-8904, Japan}

\author{Yasunobu Nakamura}
\affiliation{Research Center for Advanced Science and Technology~(RCAST), The University of Tokyo, 4-6-1 Komaba, Meguro-ku,
	Tokyo 153-8904, Japan}
\affiliation{Center for Emergent Matter Science (CEMS), RIKEN, Wako, Saitama 351-0198, Japan}

\date{\today}

\begin{abstract}
Whispering gallery modes (WGMs), circulating modes near the surface of a spheroidal material, have been known  to exhibit high quality factors for both acoustic and electromagnetic waves.  Here, we report  an electro-optomechanical system, where the overlapping WGMs of acoustic and optical waves along the equator of a dielectric sphere strongly couple to each other.   The triple-resonance phase-matching condition provides a large enhancement of the Brillouin scattering only in a single sideband, and conversion from the input radio-frequency signal exciting the acoustic mode to the output optical signal is observed.     
\end{abstract} 


\maketitle

A whispering gallery mode~(WGM), first modeled by Lord Rayleigh for the acoustic waves swirling around a spherical dome, has led to a wide range of developments in the field of optics.  A recent development in the surface preparation techniques resulted in an ultrahigh-Q microresonators exceeding $Q=10^{10}$ \cite{Grudinin2006}. Furthermore, a small mode volume of the WGM has significantly contributed to the progress, such as microlasers~\cite{Sandoghdar1996, Spillane2002}, nonlinear optics~\cite{Braginsky1989, Carmon2007}, and cavity quantum electrodynamics~\cite{Vernooy1998, Aoki2006}. Optical WGMs can also couple coherently  to phonons in the host medium by radiation pressure and photostriction.  WGM-based optomechanical systems have been used for the demonstrations of the radiation-pressure-induced parametric-oscillation instability~\cite{Kippenberg2005}, Brillouin lasing~\cite{Grudinin2009}, optomechanically-induced transparency~\cite{Weis2010}, and cavity cooling near the quantum ground state~\cite{Verhagen2012}. Among different mechanical modes, Rayleigh-type surface acoustic wave~(SAW) modes also propagate along the equator of the spheroid. Thermoelastic excitation of these modes were first observed~\cite{Yamanaka2000}, and optomechanics of the acoustic WGM~\footnote{Some authors classify the circulating acoustic modes by the radial mode number. The modes accompanied with the lowest radial mode number are referred as the Rayleigh-wave modes, while the rest of the higher-radial modes are referred as the WGMs~\cite{Shui1988}. We refer all the circulating acoustic modes as acoustic WGMs.} is theoretically studied~\cite{Matsko2009}. Realization of stimulated acoustic WGM excitation~\cite{Bahl2011} and Brillouin cooling~\cite{Bahl2012} are also reported.  In these experimental studies, only thermal or optically stimulated phonons have been exploited.

One of the anticipated applications of such photonic hybrid systems is coherent conversion of the electromagnetic waves at various energy scales \cite{Regal2011, Safavi-Naeini2019, lambert2019}.  A broadband and efficient photon converter can be used as a powerful tool such as a quantum-limited detector of low-frequency signals and a quantum transducer.  High-Q optical WGMs have been used for electro-optic modulators~(EOM).  An EOM with a lithium niobate~(LN) WGM resonator combined with a three-dimensional microwave dielectric cavity is studied to show a photon conversion efficiency $\eta=10^{-3}$~\cite{Rueda2016}. A similar EOM with a lumped-element superconducting resonator with an aluminum-nitride optical WGM has shown an internal conversion efficiency of  25.9\%~\cite{Fan2018}. Electro-optomechanical photon converters, including a high-stress membrane~\cite{Higginbotham2018},  photonic-phononic crystals~\cite{Vainsencher2016, Forsch2020, Jiang2019}, and surface acoustic waves~\cite{shao2019}, are also studied extensively.   

In this Letter, we report an electro-optomechanical system, where WGMs for both acoustic and optical waves are  exploited. The piezoelectricity of a dielectric sphere allows direct excitation of a high-Q acoustic WGM at radiowave frequency~(RF). An optical WGM is simultaneously excited and a high conversion ratio between the input RF signal and scattered optical signal is observed.  This unique geometry allows an phase-matching between the acoustic and the two optical WGMs with a large difference in the azimuthal mode number.  Additionally, temperature tuning allows to fulfill the triple-resonance condition between an acoustic and two optical WGMs, resulting in stark discrimination between the Stokes (ST) and anti-Stokes (AS) scattering processes. 

\begin{figure}[]
	\includegraphics[width=8.5cm]{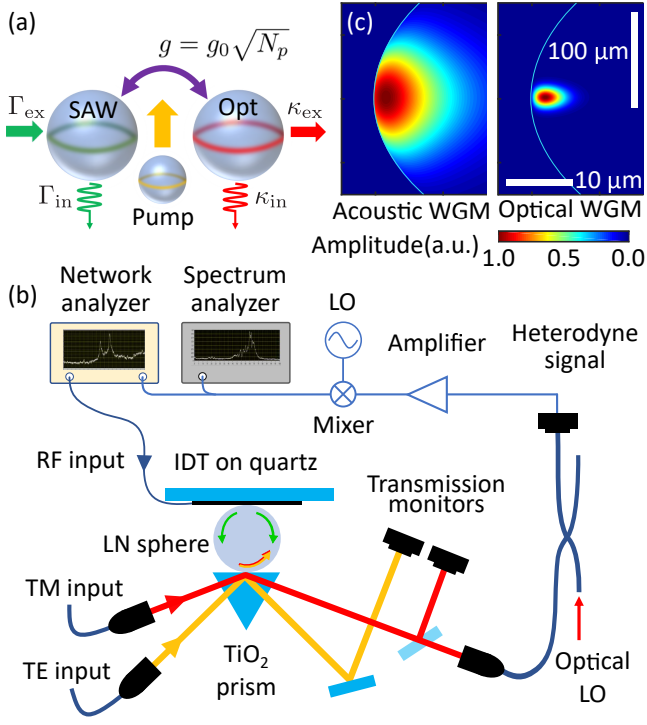}\\
	\caption{(a) Schematic architecture of the RF-to-optical conversion.  An RF signal excites the SAW mode through $\Gamma_\mathrm{ex}$.  The excited phonons are converted to the signal optical mode and extracted via $\kappa_\mathrm{ex}$.  The optical field in the pump mode enhances the coupling strength by $\sqrt{N_p}$. Note that in reality all the WGMs are on the same sphere. (b) Experimental setup:  An RF signal from a network analyzer is input to the interdigitated transducer~(IDT) patterned on a quartz substrate to excite the acoustic WGM.  Optical inputs are introduced to the sphere through a rutile (TiO$_2$) prism.  The output optical signals for the TE and TM modes are monitored by different photodiodes.  A part of the TM mode output is mixed with an optical local oscillator~(LO) for the heterodyne detection of the output signal.  The heterodyne signal is analyzed with the network analyzer and a spectrum analyzer. (c) Calculated acoustic displacement~(left) and electric field~(right) profiles for the acoustic and optical WGMs, respectively~\cite{Supplementary}.  The ratio between the vertical and horizontal scales is altered for visibility. The difference of the spatial profiles between the TE and TM optical modes is negligible. }\label{fig1}
\end{figure}
 
The architecture of the RF-to-optical conversion is illustrated in Fig.~\ref{fig1}(a). An RF signal from the external circuit excites the acoustic WGM~(green), through the piezoelectric effect, with an external coupling rate $\Gamma_\mathrm{ex}$. The resulting SAW excitation scatters the incoming photons in the pump optical mode~(orange) to the signal optical mode~(red) via photoelastic coupling $g=g_0\sqrt{N_p}$, where $g_0$ and $N_p$ are the single photon coupling strength and the intra-cavity pump-photon number, respectively.  The signal photon, then, decays out of the cavity with an external coupling $\kappa_{\mathrm{ex}}$.  As described in details in the Supplementary Material~(SM) \cite{Supplementary}, the photoelastic coupling strength can be derived as
\begin{equation}
g_0 = \sum_{i,j,k,l}\sqrt{\frac{\hbar\omega_p\omega_s}{8\rho\Omega}}\int \! d\mathbf{r} \, n_pn_sp_{ijkl} \mathcal{E}_{p i}{(\vec{\mathbf{r}})}\mathcal{E}_{s j}^*(\vec{\mathbf{r}})\nabla_l U_k, \label{eq1}
\end{equation}
where the integral is taken over the entire sample volume, $p_{ijkl}$ is the photoelastic constant, and $\mathcal{E}_{p i}$ and $\mathcal{E}_{s j}$ are the electric field amplitudes of the pump and signal optical modes with the refractive indices $n_p$ and $n_s$ and polarization direction $i$ and $j$, respectively.   The displacement amplitude of the SAW in the direction $k$ is denoted as $U_k$. $\omega_p$, $\omega_s$, and $\Omega$ are the frequencies of the optical pump, optical signal and SAW modes, respectively. $\rho$ and $\hbar$ are the material density and Planck constant divided by $2\pi$, respectively.  

Efficient conversion requires a good mode overlap of the three WGMs and the phase-matching of these traveling waves along the equator on the sphere~\cite{Matsko2009}.  In general, there are two approaches to fulfill the phase-matching condition, (i)~using two optical modes of different spatial profiles, and (ii)~using dielectric birefringence, which results in different refractive indices for two optical modes.  In our experiment, as discussed in SM~\cite{Supplementary}, it is convenient and effective to choose the latter phase-matching scheme.

For the co-propagating SAW and two optical modes, the phase-matching condition is provided as the momentum and energy conservation, reading,
\begin{equation}
k_s(\omega_s=\omega_p\pm\Omega) = k_p(\omega_p)\pm K(\Omega), \label{eq2}
\end{equation}
where $k_p$, $k_s$, and $K$ are the wavenumbers of the optical pump, optical signal and SAW modes, respectively. The plus and minus signs in Eq.~(\ref{eq2}) refer to the AS and ST processes, respectively.  In a birefringent material with $n_s \neq n_p$, the phase-matching condition sets the SAW frequency,   
\begin{eqnarray}
\Omega&=&\pm \frac{(n_s-n_p)v_{\mathrm{SAW}}\omega_p}{c-n_sv_{\mathrm{SAW}}} \nonumber   \\
&\simeq& \pm(n_s-n_p)\frac{v_{\mathrm{SAW}}}{c}\omega_p, \label{eq3} 
\end{eqnarray}
where $v_{\mathrm{SAW}}$  is the velocity of the SAW and $c$ is the speed of light in vacuum \cite{Supplementary}. For a given set of $n_p$ and $n_s$, one of the SAW frequencies is negative and non-physical.  This discrimination is often referred to as phase-matching-induced symmetry breaking~\cite{Otterstrom2018b} and plays a key role in our system.  In order for the discrete set of WGMs to fulfill the phase-matching condition above, an extra constraint is enforced on the  respective azimuthal mode numbers, $M$, $m_p$, and $m_s$, for the SAW mode and the pump and signal optical modes.  The phase-matching condition with the triple-resonance now reads 

\begin{eqnarray}
\Omega &=& |\omega_p-\omega_s|, \label{eq4} \\
M&=&|m_p-m_s|, \label{eq5}
\end{eqnarray} 
which amounts to the energy and angular momentum conservation of the entire WGM system \cite{Supplementary}.
   
A sketch of the experimental setup is shown in Fig.~\ref{fig1}(b).  Our sample is a monocrystalline LN sphere with a diameter of 3.3$~$mm.  These samples have been used previously for the demonstration of the high-Q Rayleigh-type SAW mode~\cite{Akao2004}.   For maximizing the coupling strength, we use the fundamental radial and polar modes of the SAW and two optical modes, as shown in Fig.~\ref{fig1}(c).  
The cross-sectional spatial profile of the SAW mode is large and nearly uniform in the extent of the optical modes.   The interdigitated transducer~(IDT) for the SAW excitation is made of aluminum on a quartz substrate, with the single electrode width, as well as the spacing in between electrodes, of 3.5~$\mu$m.  The IDT excites both co- and counter-propagating SAW modes with respect to the optical pump.  The SAW spectra obtained in a $S_{11}$ measurement is shown in Fig.~\ref{fig2}(a).    

\begin{figure}[]
	\includegraphics[width=8.5cm]{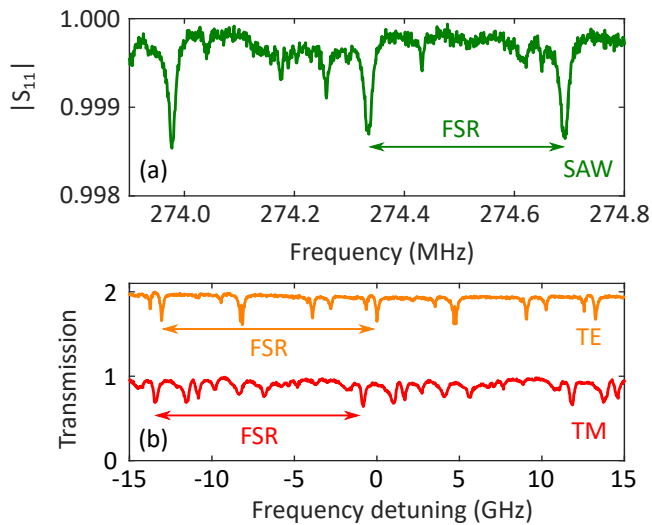}\\
	\caption{WGM spectra of SAW and optical modes.  (a)~Radio-frequency reflection spectrum of the SAW resonator.  (b)~Optical transmission spectra of the TE mode (orange) and TM mode (red). The TE mode spectrum is offset vertically by 1.  Horizontal arrows show the free spectral ranges~(FSR) of each WGM mode, $\mathrm{FSR_{SAW}} = 358$~kHz, $\mathrm{FSR_{TE}}=13.0$~GHz, and $\mathrm{FSR_{TM}}=12.6$~GHz. Dips at the edges of the horizontal arrows are the WGMs used in the experiment. The frequency detuning in the horizontal axis is defined from the TE mode used as a pump mode in the experiment. }  \label{fig2}
\end{figure} 

Next, we couple lasers into the LN sphere with a  rutile (TiO$_2$) prism.   Due to the birefringences in LN ($n_s=2.233$ and $n_p=2.156$) and TiO$_2$, the ouput beams of the optical WGMs are naturally separated for the TE and TM modes.  We perform optical WGM spectroscopy for both TE and TM polarizations simultaneously.  The obtained spectra are shown in Fig.~\ref{fig2}(b).   For the sample used in the transduction experiment, we measure $\Gamma_{\mathrm{in}}/2\pi$ = 49.7~kHz and $\Gamma_{\mathrm{ex}}/2\pi$ = 55~Hz for the SAW mode, and $\kappa_{\mathrm{in}}/2\pi$ = 80.4~MHz and $\kappa_{\mathrm{ex}}/2\pi$ = 13.5~MHz for the signal optical mode~(TM mode). For the pump optical mode~(TE mode), we measure the internal and external coupling rate $\gamma_{\mathrm{in}}/2\pi$ = 9.3~MHz and $\gamma_{\mathrm{ex}}/2\pi$ = 66.8~MHz. 
        
For the transduction experiment, we first determine the laser wavelengths and the sample temperature to attain the triple-resonance condition, where the TE and TM modes are separated by the calculated SAW frequency $\Omega/2\pi\sim 276$~MHz~[Fig.~\ref{fig3}(a)].  We observe two modes crossing each other at $T=24.3$~$^\circ$C  at $\lambda_{\mathrm{opt}}=1035.2$~nm. To look for the transduction signal, we use the TE mode as an optical pump, and  monitor the TM output signal with a photodiode, while exciting the SAW resonance with an RF signal.  We perform two dimensional scan of the RF input signal and the sample temperature.  An example of the observed $S_{11}$ signal exhibiting the acoustic WGM frequencies and the transduction signal spectrum integrated for all the temperatures are shown in Figs.~3(b) and 3(c), respectively.  We observe a transduction signal at SAW resonance $\Omega/2\pi=274.3$~MHz, with the estimated azimuthal number of $M=768$.  Small unexpected signals at 273.6 and 274.9~MHz are observed. We presume a weakly coupled higher-index  modes of the TM WGM as the origin of the signal.       
  
\begin{figure}[]
	\includegraphics[width=8.5cm]{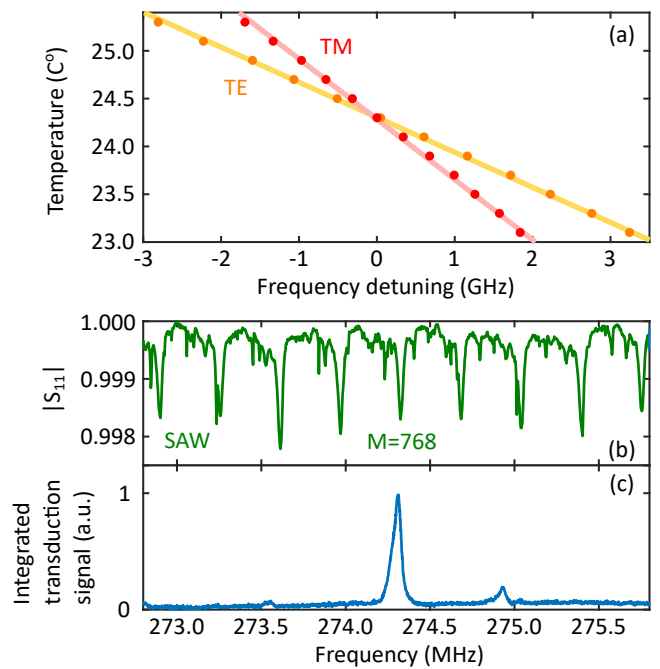}\\
	\caption{Tuning the triple-resonance condition. (a)~Frequencies of a particular set of TE~(orange) and TM~(red) modes  at different sample temperatures.  We observe the crossing temperature of 24.3~$^\circ$C.  (b)~RF reflection spectrum $S_{11}$ of the SAW modes. The deep dips observed periodically correspond to the fundamental modes of acoustic WGMs.  (c)~Integrated transduction signal for temperatures between 23.0~$^\circ$C to 26.0~$^\circ$C with the step of 0.1 $^\circ$C. as a function of the RF drive frequency.  A clear signal is seen only at at the SAW resonance with index $M=768$, indicating the triple-resonance condition. }\label{fig3}
\end{figure}

\begin{figure}[]
	\includegraphics[width=8.5cm]{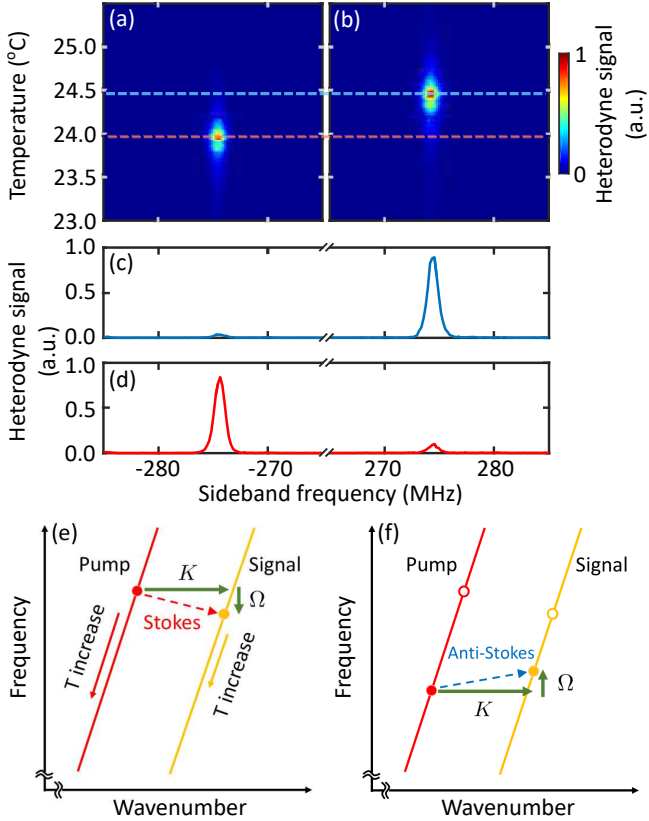}\\
	\caption{Stokes (ST) and anti-Stokes (AS) signal strength at different temperature for the SAW resonance of 274.3~MHz.  (a) and (b) shows the colormap of the heterodyne signal strength as a function of the temperature and detuning from the input laser frequency. As the sample temperature is increased, the ST sideband appears first and then the AS sideband follows.   (c) and (d) are the cross-sections of the color map at the temperatures indicated with blue and red dashed lines in (a) and (b), respectively. Linewidth of the heterodyne signals are broadened by the measurement bandwidth. 
	(e) and (f) A qualitative description of the observed signal sequence. (e) The WGM resonance, denoted as dots, first fulfill the triple-resonance phase-matching condition for the ST process.  (f) As the temperature is further increased, the AS process is phase-matched.  It is important to note the propagation direction of the SAW involved in the two processes are opposite.    }\label{fig4}
\end{figure}
After the identification of the triple-resonance condition, we lock the RF drive frequency to $M=768$ resonance and repeat the temperature scan for the heterodyne detection, where we monitor the TM output signal at the sidebands of the input laser, $\omega_p\pm \Omega$. Color maps of the observed signals are shown in Figs.~\ref{fig4}(a) and \ref{fig4}(b).  The strong signals of the ST and AS sidebands are obtained at different temperatures. Cross-sections of the color maps at the maximum signal for the AS~(blue dashed line) and ST~(red dashed line) signals are shown in Figs.~\ref{fig4}(c) and \ref{fig4}(d), respectively. A large discrimination between the ST and AS signals is observed.  A qualitative understanding of the observed signal sequence is depicted in Figs.~\ref{fig4}(e) and (f).  The signal and pump optical WGM resonances lie along the dispersion curve for each mode and move with the sample temperature.  When the sample temperature increases, the phase-matching condition [Eqs.~(\ref{eq4}) and (\ref{eq5})] for the ST process is first fulfilled~[Fig.~\ref{fig4}(e)].  As a temperature of the system further increases, the WGM resonance frequencies decrease at different rates for the two optical modes. As a result, the AS resonance condition is fulfilled at a higher temperature as shown in Fig.~\ref{fig4}(f).  Two signals originate from the Brillouin scattering by different SAW modes: Since the pump mode has a lower wavenumber, the scattering processes have to increase the wavenumber for both cases.  Thus, the AS (ST) scattering process is phase-matched only with the co-propagating (counter-propagating) SAW mode. We also performed the same scan with the TM mode being the input pump and observed the reversed sequence of the sideband appearance~\cite{Supplementary}. 

From the output signal power and the system parameters, we determine the cooperativity of the system and the RF-to-optical conversion efficiency.   
The cooperativity is defined as, $C = 4g^2_0N_p/(\kappa\Gamma)$, 
where $\Gamma$ and $\kappa$ are the total dissipation rates of the SAW mode ($\Gamma=\Gamma_{\mathrm{in}}+\Gamma_{\mathrm{ex}}$) and the signal optical mode ($\kappa=\kappa_{\mathrm{in}}+\kappa_{\mathrm{ex}}$), respectively. The intra-cavity photon number is given as $N_p=4P_{\mathrm{in}}\gamma_{\mathrm{ex}}/(\hbar\omega_p\gamma^2)$, where $\gamma=\gamma_{\mathrm{in}}+\gamma_{\mathrm{ex}}$, for  the input optical power $P_\mathrm{in}$.
As derived in the SM~\cite{Supplementary}, the conversion efficiency $\eta$ can be expressed in terms of the cooperativity $C$ in a simple form
\begin{equation}
	\eta = \frac{\kappa_\mathrm{ex} }{\kappa}\frac{\Gamma_\mathrm{ex}}{\Gamma}\frac{4C}{(1+C)^2}. \label{eq7} 
\end{equation}
We measure the maximum conversion efficiency of $\eta=1.2\times10^{-6}$, with the cooperativity $C=2.0\times10^{-3}$ at the optical input pump power of 372~$\mu$W.  The single photon coupling strength is determined as $g_0/2\pi=12$~Hz, which is lower than a numerical estimate of $g_0/2\pi=24.5$~Hz using the photoelastic constant of LN, $|p_{ijkl}| = 0.11$.     
   
As discussed in detail in the SM~\cite{Supplementary}, there is a large margin for the improvement of the system. The main bottleneck of the current system is definitely at the RF input. A geometrical mismatch of the sphere and the flat IDT resulted in the weak external coupling $\Gamma_\mathrm{ex}/2\pi=55$~Hz with $\Gamma_{\mathrm{ex}}/\Gamma$ ratio of 1/1,100.  One can fabricate airbridge IDTs directly on the sphere to enhance the external coupling while retaining the high Q-factor of both the acoustic and optical WGMs.  We estimate the external coupling enhancement of at least a factor of 10.  The IDT with an unidirectional excitation can also enhance the performance by a factor of 2. Additionally, the geometrical modification of the sample can significantly improve the performance by choosing a smaller sphere or to utilize a disk-type resonator for a further reduction of the WGM volume~\cite{Spillane2005}.     Assuming $Q_{\mathrm{opt}}=5\times10^7$ and $Q_{\mathrm{SAW}}=5\times10^4$, which we routinely observe, and the modifications above for the LN sphere with a diameter 330~$\mu$m, we expect $C\sim1$ and $\eta=6.9\times 10^{-2}$ for the  optical pump power of 3~$\mu$W. A particular importance of this system lies in the high quality factors obtained by WGMs, despite the large sample size compared to other mechanical systems. Even with a relatively large sample size, which results in a small coupling $g_0$, a large intra-cavity photon number $N_p$ in a high-Q optical WGM can compensate to attain a large $C$ with a little optical pump.  The small input power could be a large advantage in the dilution refrigerator environment, where an excess of the stray photons will be detrimental.  The triple-resonance condition, which we attained by the temperature control, could also be achieved by a pressure-induced frequency shift of the WGMs.  Further investigation with these improvement could be an important step towards a realization of a unit-efficiency coherent electro-optical conversion.

In summary, we developed an electro-optomechanical system comprises of acoustic and optical WGMs.  A direct excitation of the SAW on the WGM was realized with an external IDT.  Under the triple-resonance condition, a large enhancement of the RF-to-optical conversion was observed.          
 
\begin{acknowledgments}
The authors acknowledge H. Takahashi for the help in sample fabrication. This work was partly supported by JSPS KAKENHI (Grant No. 26220601), NICT, JST PRESTO (Grant No. JPMJPR 1429), and JST ERATO (Grant No. JPMJER1601).
\end{acknowledgments}


\appendix
\renewcommand{\thefigure}{A\arabic{figure}} 
\setcounter{figure}{0}
\renewcommand{\theequation}{A\arabic{equation}} 
\setcounter{equation}{0}
\renewcommand{\thetable}{A\arabic{table}} 
\setcounter{table}{0}

\maketitle
\onecolumngrid
\section{Experimental Setup}
The LN sphere used in the experiment is originally developed for the SAW-based gas sensor~\cite{Yamanaka2006} and gas chromatography~\cite{Iwaya2012} and has a high sphericity and surface quality.  Both the acoustic and optical WGMs of interest lie along the equator of an LN sphere with the $c$-axis of the crystal being the polar axis.  A rough estimate of the $c$-axis of the lithium niobate (LN) sphere is pre-performed before mounting with a conoscopic interferometric method. The LN sphere is glued on a stainless rod and is mounted on a sample holder consisting of a goniometer and three-axis translation stages along with a peltier module for the temperature control.  After mounging the sample, the $c$-axis orientaion is further adjusted by sending a polarized laser through the center of the sphere and minimizing the polarization rotation on the transmitted laser.

After the sphere orientation adjustment, an interdigitated transducer~(IDT) for the SAW excitation is prepared for the SAW excitation.  An IDT made of aluminum is pattered on a quartz substrate.  The single-electrode width, as well as the spacing between electrodes, is 3.5~$\mu$m, designed for the SAW frequency of approximately $\Omega/2\pi=270$~MHz.  The IDT substrate is mounted and wire-bonded to an IDT holder with a SMA connector, which is connected to a network analyzer. When the IDT is placed near the sphere surface, the SAW signal appears on the reflection spectrum. IDT misalignment causes the excitation of the higher-index (polar) modes, showing up as sub-peaks of the main SAW signal.  The alignment, including the tilt and vertical position of the IDT, is adjusted to minimize the higher-index mode signals.  

\begin{figure}[h]
  	\includegraphics[width=15cm]{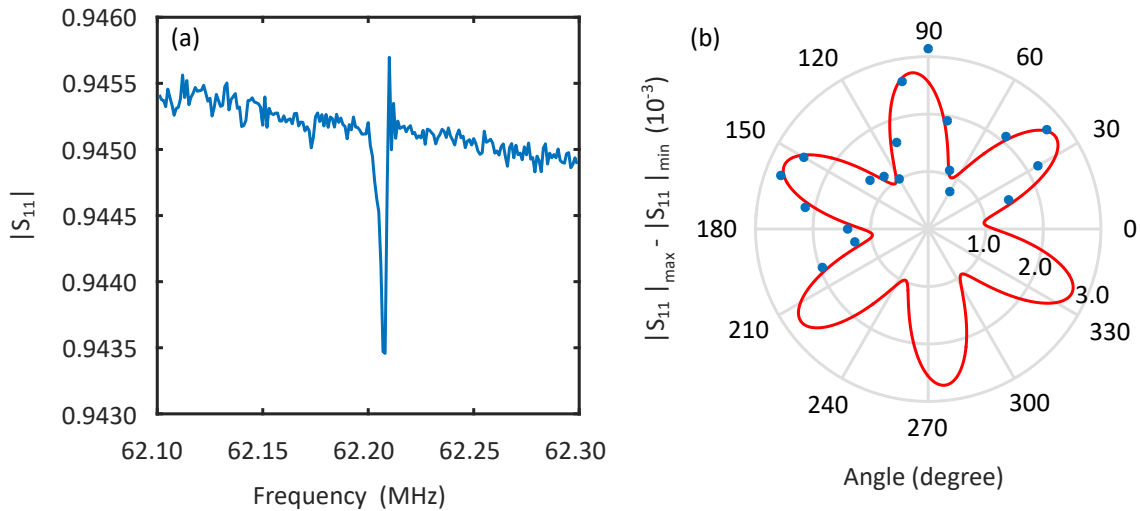}\\
  	\caption{IDT alignment. (a) $S_{11}$ signal obtained during the IDT alignment.  The external coupling of the acoustic WGM can be obtained from the magnitude of the dip in the signal.  (b) $S_{11}$ dip magnitude as a function of the crystal rotation along the $c$-axis.  The IDT is placed at one of the 6-fold external coupling maxima.  }\label{figS1}
  \end{figure}

The piezoelectric coupling has 6-fold symmetry with respect to the rotation along the $c$-axis.  An example of the reflection signal spectrum and the signal strength with respect to the sphere rotation angle are shown in Fig.~\ref{figS1}. We optimize the sphere rotation angle by maximizing the depth of the SAW resonance signal.   The maximum Q-factor obtained for the SAW mode is $7.3\times10^4$, which is close to the air damping limit previously reported for LN \cite{1974Li,1975Li}.  We also observe the free spectral range (FSR) of 357.50~kHz, from which we determine the SAW velocity $v_{\mathrm{SAW}}=3706$~m/s.
    
Next, we couple lasers into the LN sphere by a prism.  The refractive indices of TE and TM modes in the sphere are large ($n_\mathrm{TE}=2.156$ and $n_\mathrm{TM}=2.233$).  Thus, we have only a few options for the prism material. The prism we use is made of rutile (TiO$_2$), with 60-60-60 degree cut, mounted on a translation stage.  The input and output beams for the WGMs (TE or TM) are separated in angle respectively, and two photodetectors are used to monitor the output signals from the WGMs.  

Different optical WGMs can be identified by varying the input laser incident angles and monitoring the output angles and the beam shapes \cite{Schunk2014}.  Following the previous studies on WGM, we also describe the mode assignments  with  radial, polar, and azimuthal quantum numbers, $q$, $l$, and $m$, respectively. The input lasers are adjusted to optimize the coupling to the optical WGMs with $q$=1 and $l=m$.  During the transduction experiment with the TE mode being the optical pump mode, we monitor the TM output signal on the photodetector.  

A part of the TM output beam is sent to a polarization maintaining fiber, mixed with an optical local oscillator signal, which is derived from the same laser used as a pump beam, but offseted by $-80$~MHz with an AOM, for the heterodyne detection.  The heterodyne detection with an offset laser allows us to discriminate the Stokes and anti-Sotkes signals on the spectrum analyzer.  

\section{RF and Optical Calibration}
\begin{figure}[h]
	\includegraphics[width=15cm]{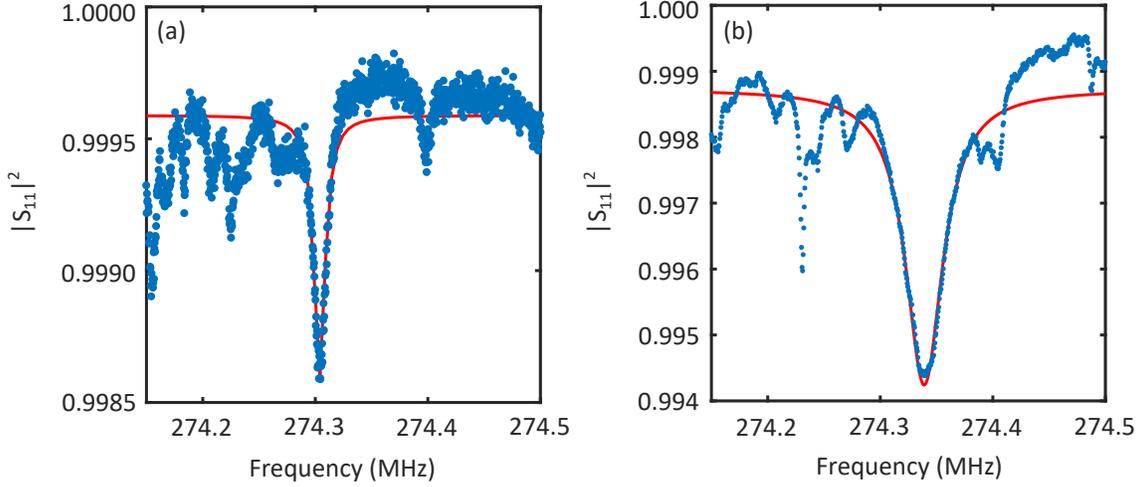}\\
	\caption{SAW linewidth measurement.  (a) SAW reflection signal with the IDT only, (b) SAW reflection signal when the optical prism is introduced. The red lines are the Lorentzian fittings. }\label{figS2}
\end{figure}

RF loss in the cable, connectors, and IDT are measured and calibrated by monitoring the reflection spectra ($S_{11}$) away from the SAW resonance. Placing an optical prism introduces an extra damping of the SAW, and hence increases the internal loss.  As mentioned in the main text, the IDT--SAW coupling is largely undercoupled and the linewidth of the SAW resonance is nearly the intrinsic linewidth of the SAW.  SAW resonance profiles with the IDT only (without an optical prism) and with the optical prism, as in the transduction experiment, are shown in Fig.~\ref{figS2}.  We measure the loaded Q-factor of $Q=1.85\times 10^4$ and $5.5\times10^3$, respectively, with the corresponding SAW decay rate of $\Gamma/2\pi=14.8$~kHz and 49.7~kHz.  

External coupling rate $\Gamma_{\mathrm{ex}}/2\pi=55$ Hz is obtained from fitting the depth of the $S_{11}$ signal and the total loss rate $\Gamma$.  The weak external coupling can be understood with the following formula, given as a function of the resonator parameters \cite{Morgan2007},
\begin{equation}
\Gamma_{\mathrm{ex}}=\frac{5.74 v_{\mathrm{SAW}} Z_c \epsilon K^2 W N^2\Omega}{L },
\end{equation}
where $Z_c$ is the impedance of the IDT drive port, $\epsilon$ and $K^2$ are the dielectic constant and piezoelectric coupling strength of LN, respectively. $W$, $L$, and $N$ are the width and length of the SAW resonator, and the number of the pairs of IDT electrodes, respectively.  In an acoustic WGM, a tight confinement (small $W$) along with a relatively large cavity length $L$ results in a weak external coupling.  

With our device parameters~($\lambda_\mathrm{SAW} = 14$~$\mu$m, $Z=50$~$\Omega$, $\epsilon=4.6\times10^{-4}$, $K^2=0.02$, $W=60$~$\mu$m, $L=10$~mm, $\Omega/2\pi=274$~MHz), we calculate a single-pair IDT external coupling of $\Gamma_{\mathrm{ex}}/2\pi=98$~Hz.   This calculation is consistent with the measured external coupling strength and the large geometrical mismatch between the flat IDT substrate and the spherical LN sample.  We presume only a pair of IDT electrodes is effectively participating in the excitation and also the air gap between the IDT and the sample diminishes the coupling further.    

For the optical calibration, the optical power at various location of the the individual paths for the TE and TM polarization is measured to incorporate the losses. The rutile prism used for the optical WGM coupling is uncoated and a surface reflection could be as large as 20\%.  We measure the power differences between the input and output of the prism for both polarizations to calibrate for a large reflection. All the photodetectors are calibrated by inserting a laser with a known power measured with an optical power meter.  Moving the prism with respect to the sphere allows to vary the external optical coupling.  We measure the internal optical loss rate $\kappa_\mathrm{in}$ by either, (i)~moving the prism away and measure the largely undercoupled optical linewidth, or (ii)~identifying the critical-coupling condition by moving the prism to find the largest absorption on the transmission signal.  

The maximum Q-factor obtained for the optical WGMs is $1.08\times10^8$, near the absorption limit of the material \cite{Savchenkov2004}.  The internal optical decay rate $\gamma_\mathrm{in}/2\pi=9.3$~MHz and $\kappa_\mathrm{in}/2\pi=80.4$~MHz are obtained for the TE and TM modes, respectively.  The second procedure mentioned also helps identifying the spatial matching of the input laser by observing the depth of the absorption at the critical-coupling condition.   We determined the spatial mode matching between the input laser and the optical WGM to be 0.33 for our experiment.      

\section{Phase-Matching Condition and Triple-Resonances}
For the co-propagating SAW and pump optical modes, the phase-matching conditions (momentum and energy conservation) for the Brillouin scattering require,
\begin{eqnarray}
k_s&=& k_p\pm K \label{pm01}, \\
\omega_s&=&\omega_p\pm\Omega.  \label{pm02} 
\end{eqnarray}
The plus and minus signs corresponds to the anti-Stokes and Stokes processes, respectively.  Rewriting the momentum conservation in terms of frequencies using the dispersion relations, $\frac{\omega}{k}=\frac{c}{n}$ for the optical modes and $\frac{\Omega}{K}=v_{\mathrm{SAW}}$ for the SAW mode, and substituting Eq.~(\ref{pm02}), the phase-matched SAW frequency reads,
\begin{equation}
\Omega=\pm \frac{(n_s-n_p)v_{\mathrm{SAW}}\omega_p}{c-n_sv_{\mathrm{SAW}}}. \label{pm03}
\end{equation}
Similarly, for the counter-propagating SAW and pump optical modes, the phase-matched SAW frequency is gives as,
\begin{equation}
\Omega=\mp \frac{(n_s-n_p)v_{\mathrm{SAW}}\omega_p}{c+n_sv_{\mathrm{SAW}}}. \label{pm04}
\end{equation}

Once the pump optical frequency $\omega_p$ is set, the material constants solely determines the phase-matching SAW frequency.  For the current experiment, we calculate $\Omega/2\pi=276$~MHz with the refractive indices $n_s=2.233$ and $n_p=2.156$, the SAW velocity $v_\mathrm{SAW}=3706$~m/s, and the pump optical frequency $\omega_p/2\pi=2.9\times10^{14}$~Hz.  The negative SAW frequency, for a given set of refractive indices ($n_s$ and $n_p$), indicates a non-physical solution.  For our experiment with $n_s>n_p$, only the anti-Stokes (Stokes) process for the co-propagating (counter-propagating) SAW and optical pump is allowed.    
Additionally, the WGM resonance condition also requires, 
\begin{equation}
m_i\lambda_i = 2\pi R, \label{pm05}
\end{equation}
where $m_i=m_s$, $m_p$, $M$ are the azimuthal mode numbers and $\lambda_i=\lambda_s/n_s$, $\lambda_p/n_p$, $\lambda_{\mathrm{SAW}}$ are the wavelengths, of the signal optical WGM, the pump optical WGM, and the acoustic WGM, respectively.  Substituting Eq.~(\ref{pm05}) to Eq.~(\ref{pm01}), the phase-matching with the triple-resonance condition now reads,
\begin{equation}
M = |m_p-m_s|. \label{pm06}
\end{equation}
For the current experiment with the TE (TM) mode being the pump (signal) mode, the birefringence of LN defines the relative magnitude of wavenumbers, 
\begin{equation}
k_s>k_p\gg K. \label{pm07}
\end{equation}

\begin{figure}[]
	\includegraphics[width=5.0cm]{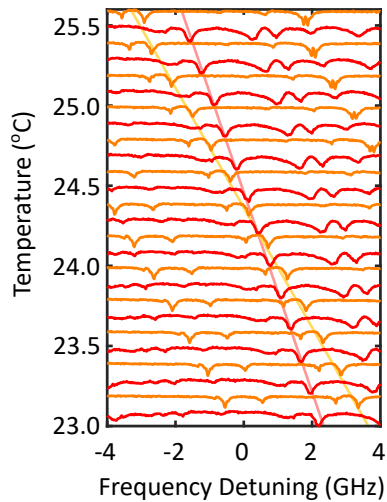}\\
	\caption{Tuning the triple-resonance condition. TE~(orange) and TM~(red) mode spectra at different sample temperatures. Both spectra are obtained at the same temperature with an increment of 0.2 $^\circ$C and TM signal is offset by $+0.1$ $^\circ$C to avoid overlap. Light-orange and light-red straight lines are guides to the eye to indicate the positions of the fundamental TE and TM resonances. We observe the crossing temperature of 24.3~$^\circ$C.   }\label{figS3}
\end{figure}

As described in the main text, we perform the transduction experiment with the input laser, pump laser, locked on either the TE or TM mode.   In order to fulfill the triple-resonance condition, $\Omega=|\omega_p-\omega_s|$, we control the temperature of the LN sphere to tune the relative frequency difference between the pump and signal optical WGMs.  Fig.~\ref{figS3} shows the temperature dependence of the actual TE and TM spectra.  In the main text, only the resonance frequencies with respect to the sample temperatures are plotted in Fig.~3(a).  We obtain the temperature dependence of the resonance shift of  $-2.76$~GHz/K and $-1.77$~GHz/K for the TE and the TM modes, respectively, with a differential shift of $0.99$~GHz/K. The resonance frequencies for the two WGMs coincide at approximately 24.3~$^\circ$C.  

\begin{figure}[h]
	\includegraphics[width=15cm]{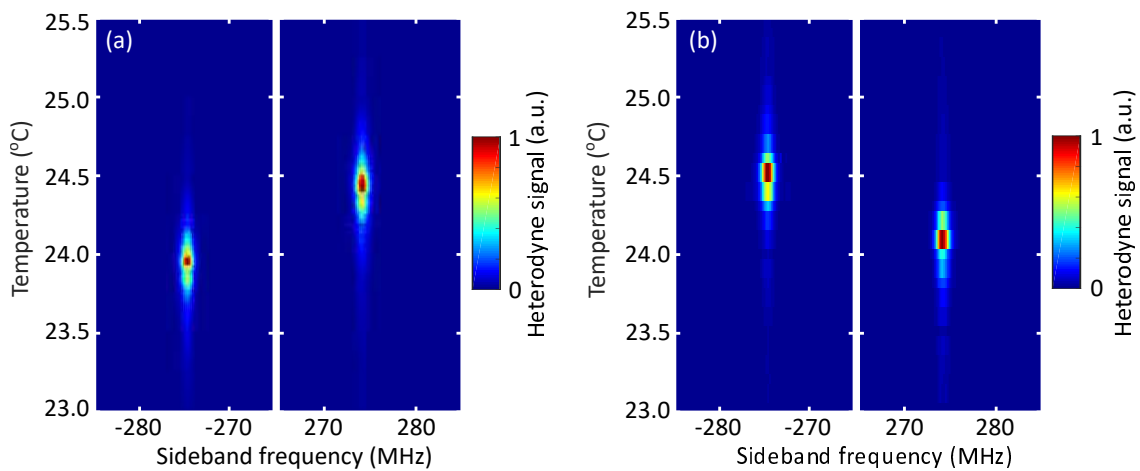}\\
	\caption{Heterodyne signal with different optical pump modes.  (a) TE mode as the pump mode. (b) TM mode as the pump mode. The two data sets are taken with different temperature resolutions. }\label{figS4}
\end{figure}

Colormaps of the heterodyne signal at the blue (red) sideband, corresponding to anti-Stokes (Stokes) process are shown in Fig.~\ref{figS4}.   Fig.~\ref{figS4}(a) is the same data as Fig.~4(a) and 4(b) in the main text.  In the experiment, the TM mode resides on the red side of the TE mode at lower temperature and shift to the blue side near 24.3~$^\circ$C as the temperature is increased~[Fig.~3(a) in the main text].  With the TE mode being the pump [Fig.~\ref{figS4}(a)], the heterodyne signal first appears as the Stokes signal and the anti-Stokes signal follows as the temperature is increased.  For the TM mode being the pump [Fig.~\ref{figS4}(b)], the role of two optical WGMs switches and the anti-Stokes signal is observed at a lower temperature. 

\section{Photoelastic Effect}
Photoelastic effect can be described as a change of the system energy due to the variation of the permittivity $\Delta\epsilon_{ij}$ resulting from the elastic deformation, provided as,

\begin{eqnarray}
\Delta U &=& \sum_{i,j}\frac{\epsilon_0}{2}\int \! d\mathbf{r} \, \Delta\epsilon_{ij} E_{1 i}{(\vec{\mathbf{r}})}E_{2 j}(\vec{\mathbf{r}}) \nonumber \\
&=& \sum_{i,j,k,l}-\frac{\epsilon_0}{2}p_{ijkl}\int \! d\mathbf{r} \, n_i^2n_j^2E_{1i}{(\vec{\mathbf{r}})}E_{2j}(\vec{\mathbf{r}})\frac{\partial u_k(\vec{\mathbf{r}})}{\partial x_l}, \label{eqS1} 
\end{eqnarray}
where the integral is taken over the entire sample volume.   $p_{ijkl}$ is the photoelastic constant, $E_{1i}$ and $E_{2j}$ are the electric field amplitude of modes 1 and 2 with polarization directions $i$ and $j$, and corresponding refractive indices $n_i$ and $n_j$, respectively.   $u_k$ is the displacement amplitude of the elastic wave in the direction $k$. $\epsilon_0$ and $\hbar$ are vacuum permittivity and Planck constant divided by $2\pi$, respectively.  Defining the operators for the electric fields and mechanical displacement as

\begin{eqnarray}
E_{\gamma i}(\vec{\mathbf{r}}) &=& \sqrt{\frac{\hbar\omega_\gamma}{2\epsilon_i}}\left[\mathcal{E}_{\gamma i}(\vec{\mathbf{r}})a_\gamma(t)+\mathcal{E}_{\gamma i}^*(\vec{\mathbf{r}})a_\gamma^\dagger(t)\right] \label{eqS2}\\
u_k(\vec{\mathbf{r}}) &=& \sqrt{\frac{\hbar}{2\rho\Omega}}\left[U_k(\vec{\mathbf{r}})b(t)+U_k^*(\vec{\mathbf{r}})b^\dagger(t)\right],\label{eqS3} 
\end{eqnarray}
where $a_\gamma$ and $b$ are the creation operators of the optical mode $\gamma$ and the mechanical mode.  $\omega_\gamma$ and $\epsilon_i$ are the angular frequency of the optical mode $\gamma$ and the permittivity of the polarization direction in $i$, respectively.  $\rho$ and $\Omega$ are the material density and the mechanical angular frequency, respectively.  The normalization constants for the electric fields and mechanical displacement follows normalization conditions,
\begin{eqnarray}
\sum_i\int \! d\mathbf{r} \, \mathcal{E}_{\gamma i}{(\vec{\mathbf{r}})}\mathcal{E}_{\gamma i}^*(\vec{\mathbf{r}}) &=&1 \\
\sum_k\int \! d\mathbf{r} \, U_k(\vec{\mathbf{r}})U_k(\vec{\mathbf{r}})^* &=&1. 
\end{eqnarray}

In the current experiment, one of the optical mode is a pump mode with a large field amplitude, allowing a replacement
\begin{equation}
\hat{a}_p(t) \rightarrow \alpha_p(t)=\sqrt{N_p}\,e^{i\omega_p t},
\end{equation}
where $\alpha_p$ is a $c$-number whose magnitude is equal to a square root of the intra-cavity photon number $N_p$ in the mode.    Substituting Eqs.~(\ref{eqS2}) and (\ref{eqS3}) to Eq.~(\ref{eqS1}),  eliminating the counter rotating optical terms ($\hat{a}_s\hat{a}_p$ and $\hat{a}_s^\dagger \hat{a}_p^\dagger$), we obtain an interaction Hamiltonian,
\begin{equation}
H_\mathrm{int} =-\hbar g_0\sqrt{N_p}\,(\hat{a}_s e^{-i\omega_p t} + \hat{a}_s^\dagger e^{i\omega_p t})(\hat{b}+\hat{b}^\dagger),\label{eqS15}\\
\end{equation}
where a single photon coupling strength $g_0$ is
\begin{equation}
g_0 = \sum_{i,j,k,l}\sqrt{\frac{\hbar\omega_p\omega_s}{8\rho\Omega}}\int \! d\mathbf{r} \, n_pn_sp_{ijkl} \mathcal{E}_{p i}{(\vec{\mathbf{r}})}\mathcal{E}_{s j}^*(\vec{\mathbf{r}})\frac{\partial U_k(\vec{\mathbf{r}})}{\partial x_l}.  \label{eqS16}
\end{equation}

One can calculate the single photon coupling strength above, using the approximate spatial distribution of the  optical ($\mathcal{E}_{\gamma i}$) and acoustic ($U_k$) WGMs in a sphere, given by~\cite{Matsko2009, Breunig2013},
\begin{eqnarray}
\mathcal{E}_{\gamma i}(r,\theta,\phi)&=& \mathcal{E}_{\gamma i0} \mathrm{Ai}\!\left(\frac{2^{1/3}m^{2/3}(R-r)}{R}-\zeta_q\right)e^{-m\theta^2/2}H_p(\sqrt{m}\theta)\,e^{im\phi} \\
U_k(r,\theta,\phi) &=& U_{k0}\left(e^{-K(R-r)}-1.84e^{-0.3K(R-r)}\right) e^{-M\theta^2/2}H_p\!\left(\sqrt{M}\theta\right)e^{iM\phi},
\end{eqnarray}
where $\mathcal{E}_{\gamma i0}$ and $U_{k0}$  are the normalization constants for optical field and SAW displacement field, respectively.  $R$ is the radius of the sphere, $H_p$ and Ai are Hermite polynomial of order $p$ and airy function, respectively.  $q$, $p=l-m$ (or $p=L-M$ for SAW), $m$ (or $M$) are the radial, polar, an azimuthal mode number of the WGMs. $\zeta_q$ is the $q$-th zero of the airy function.  In the current experiment, we use fundamental radial and polar WGM modes with $q = p = 1$ ($\zeta_1\sim2.338$). As shown in Fig.~1(c) of the main text, the cross-sectional spatial profile of the SAW mode is large and nearly uniform in the extent of the optical modes. Choosing two orthogonal optical WGMs with different spatial modes, instead, results in nulling the integral in Eq.~\ref{eqS16}, which leads to a small photoelastic coupling. 

\begin{figure}[h]
	\includegraphics[width=8.5cm]{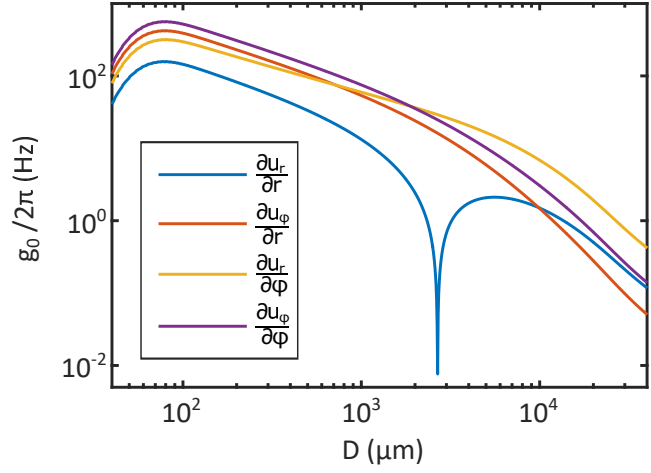}\\
	\caption{Numerical calculation of $g_0$ as a function of the sphere diameter $D$. Separate lines show the coupling strength calculated from different strain components of the SAW.}\label{figS5}
\end{figure}

Defining the $c$-axis of the LN sphere to be $z$-axis, the stain components $S_{kl}=\frac{1}{2}\left(\frac{\partial u_k}{\partial x_l}+\frac{\partial u_l}{\partial x_k}\right)$ relevant to the current experiment are $S_1$, $S_2$, and $S_6$, using contracted notation where 11 $\rightarrow$ 1; 22 $\rightarrow$ 2; 33 $\rightarrow$ 3; 23, 32 $\rightarrow 4$; 31, 13 $\rightarrow$ 5; 12, 21 $\rightarrow$ 6. Similarly, relevant photoelastic constants $p_{ijkl}$ are $p_{41}$, $p_{42}$, and $p_{56}$, also written in the contracted notation, with all of them having the same constant value $|p|=0.11$.   We numerically calculate the single photon coupling strength $g_0$ with respect to the sphere radius, and the results are shown in Fig.~\ref{figS5}.  $p_{ijkl}=0.11$ is used for the photoelastic constant.  Different color lines show the coupling strengths for different strain components denoted in the legend. A sharp dip in one of the strain component arises from nulling of the overlap integral between the optical and SAW radial modes.   We estimate $g_0/2\pi=24.5$~Hz from the biggest contribution (yellow line) for our current sample.   

\section{Conversion Efficiency}
\begin{figure}[h]
	\includegraphics[width=10cm]{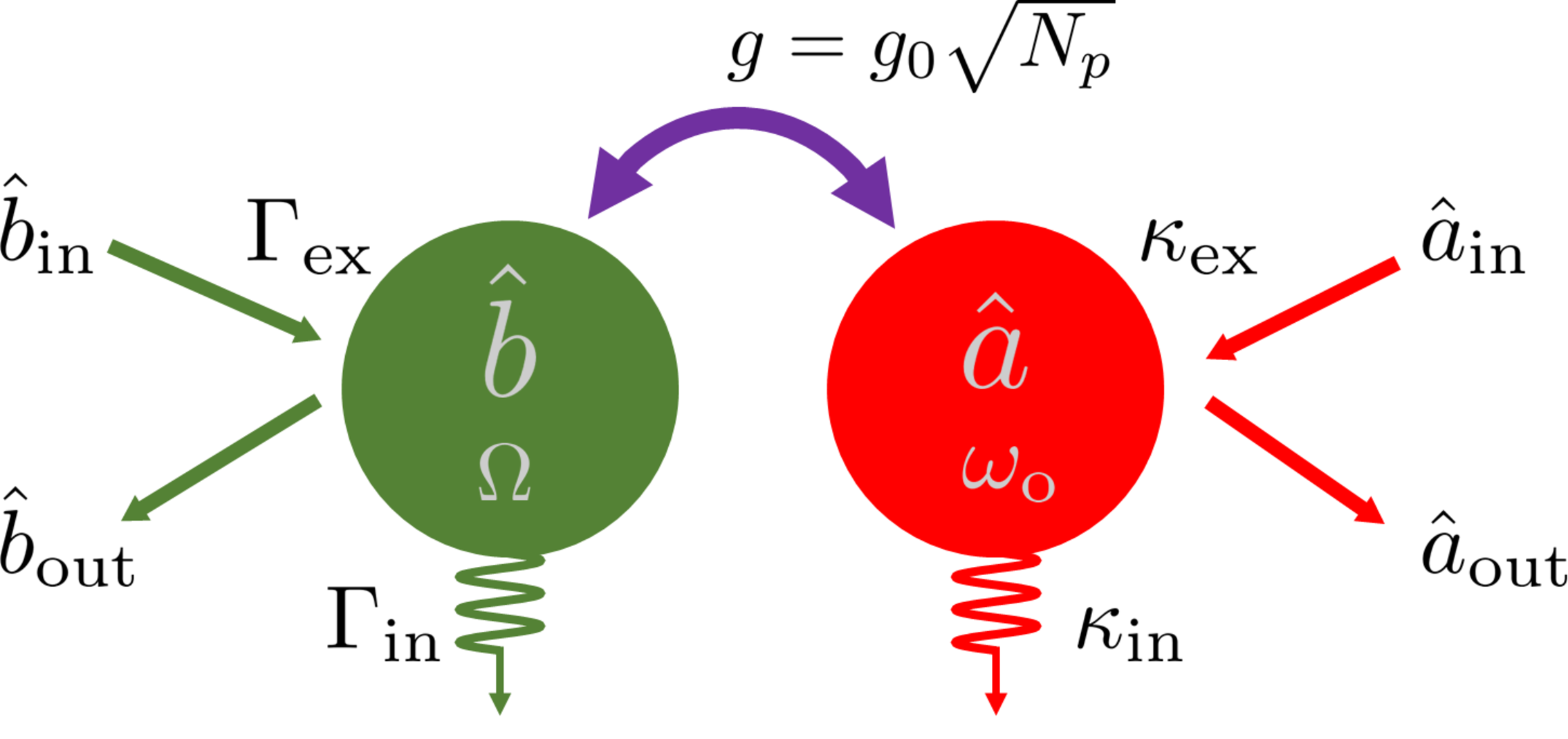}\\
	\caption{Architecture of the RF-to-optical converter.
		An optical WGM mode $\hat{a}$ and acoustic WGM $\hat{b}$ with corresponding resonance frequencies $\omega_o$ and $\Omega$, are coupled via photoelastic coupling with single photon coupling $g_0$. Each cavity is associated with the internal and external coupling rates, $\kappa_\mathrm{in}$ and $\kappa_\mathrm{ex}$ for the optical mode and  $\Gamma_\mathrm{in}$, and $\Gamma_\mathrm{ex}$ for the SAW mode, respectively.  The itinerant RF and optical modes in and out of the cavity modes are denoted as $\hat{a}_\mathrm{in}$, $\hat{a}_\mathrm{out}$, $\hat{b}_\mathrm{in}$, and $\hat{b}_\mathrm{out}$.  
		}\label{figS6}
\end{figure}

The architecture of the RF-to-optical converter is depicted in Fig.~\ref{figS6}.  Using the interaction Hamiltonian, Eq.~(\ref{eqS15}), and moving into a rotating frame of $\hbar\omega_p\hat{a_s}^\dagger\hat{a_s}$ under the rotating-wave approximation, the system Hamiltonian now reads with replacing the signal optical mode operator $\hat{a}_s$ with $\hat{a}$,
\begin{eqnarray}
H&=&\hbar\Delta\hat{a}^\dagger\hat{a}+\hbar\Omega\hat{b}\hat{b}^\dagger-\hbar g_0\sqrt{N_p}\,(\hat{a}^\dagger\hat{b}+\hat{a}^\dagger\hat{b}),
\end{eqnarray}
where $\Delta=\omega_o-\omega_p$.

According to the derivation in Ref.~\cite{Hisatomi2016}, the optical and SAW cavity modes follow the equations of motion in the Fourier domain,
\begin{eqnarray}
\hat{a}(\omega)&=&\chi_o(\omega)\left[-\sqrt{\kappa_{\mathrm{ex}}}\,\hat{a}_{\mathrm{in}}(\omega)-ig\hat{b}(\omega)\right],\\
\hat{b}(\omega)&=&\chi_m(\omega)\left[-\sqrt{\Gamma_{\mathrm{ex}}}\,\hat{b}_{\mathrm{in}}(\omega)-ig\hat{a}(\omega)\right],
\end{eqnarray}
  where the susceptibility $\chi_o(\omega)$ and $\chi_m(\omega)$ are defined as
\begin{eqnarray}
  \chi_o(\omega)&=&\frac{1}{-i(\omega-\Delta)+\frac{\kappa_{\mathrm{in}}+\kappa_{\mathrm{ex}}}{2}},\\
  \chi_m(\omega)&=&\frac{1}{-i(\omega-\Omega)+\frac{\Gamma_{\mathrm{in}}+\Gamma_{\mathrm{ex}}}{2}}.
\end{eqnarray}
We focus on the RF-to-optical conversion efficiency and solving the equations of motion with the boundary conditions
\begin{eqnarray}
\hat{a}_{\mathrm{out}}(\omega)&=&\hat{a}_{\mathrm{in}}(\omega)+\sqrt{\kappa_{\mathrm{ex}}}\,\hat{a}(\omega),\\
\hat{b}_{\mathrm{out}}(\omega)&=&\hat{b}_{\mathrm{in}}(\omega)+\sqrt{\Gamma_{\mathrm{ex}}}\,\hat{b}(\omega),
\end{eqnarray}
the RF input and optical output modes can be derived as
\begin{eqnarray}
\hat{b}_{\mathrm{in}}(\omega)&=&ig\sqrt{\frac{\kappa_{\mathrm{in}}}{\Gamma_{\mathrm{ex}}}}\,\chi_o(\omega)\hat{a}_{\mathrm{in}}(\omega)-\frac{1}{\sqrt{\Gamma_{\mathrm{ex}}}\,\chi_m(\omega)}\left[1+g^2\chi_o(\omega)\chi_m(\omega)\right]\hat{b}(\omega),\\
\hat{a}_{\mathrm{out}}(\omega)&=&\left[1-\kappa_{\mathrm{ex}}\chi_o(\omega)\right]\hat{a}_{\mathrm{in}}(\omega)-ig\sqrt{\kappa_{\mathrm{ex}}}\,\chi_o(\omega)\hat{b}(\omega).
\end{eqnarray}
We neglect the noise coming in from the optical input port $\hat{a}_{\mathrm{in}}$, a photon conversion efficiency $\eta$ is given by,

\begin{eqnarray}
\eta&=&\left|\left\langle \frac{\hat{a}_{\mathrm{out}}}{\hat{b}_{\mathrm{in}}}\right\rangle\right|^2 \\
&=&\frac{\kappa_{\mathrm{ex}}\Gamma_{\mathrm{ex}}\chi_o^2(\omega)\chi_m^2(\omega)g^2}{[1+g^2\chi_o(\omega)\chi_m(\omega)]^2}. 
\end{eqnarray}
 
 In terms of cooperativity factor $C=\frac{4g^2}{\kappa\Gamma}$, the conversion efficiency at the resonance condition $\omega_o-\omega_p=\Omega$ now reads
 
 \begin{equation}
 \eta = \frac{\kappa_{\mathrm{ex}}\Gamma_{\mathrm{ex}}}{\kappa\Gamma}\frac{4C}{(1+C)^2}.
 \end{equation}

\section{System Improvement}
The system could be miniaturized to further confine the acoustic and optical WGMs to enhance the coupling. A current polishing technologies allow to fabricate sub-millimeter crystalline microresonators with extremely high-Q.  A magnesium fluoride WGM resonator with the diameter of 700~$\mu$m and intrinsic Q-factor exceeding $10^9$ has been fabricated \cite{Wang2013}.  We discuss a improvement of the system parameter with a LN microresonator with the diameter of 330~$\mu$m, 1/10 of the size of the current LN sphere.  

We numerically calculate the photoelastic coupling $g_0/2\pi=2.0\times10^2$~Hz for this microresonator. The Q-factor of the resonator should not be decreased with the current polishing technology and is expected to remain as high as $Q=5\times 10^7$, which we routinely measure with the current system.  

One of the bottlenecks in the current system is the weak piezoelectric coupling arising from the geometrical mismatch between the flat IDT substrate and spherical sample.  Our analysis shows that only a pair of IDT electrodes participating in the SAW excitation in the current system.  IDTs have been fabricated directly on a sphere, or on a concaved substrate with a microbeads spacer \cite{Akao2004, Yamanaka2007}.  The airbridge fabrication technique can be incorporated to prepare non-contact IDTs few hundreds of nanometers above the equator of the sphere.  As the air-gap widens, to ensure not to disturb the evanescent optical wave with the IDT electrodes, the electric field inside the LN for the piezo excitation can rapidly diminishes, and many IDTs are necessary for the compensation.  We numerically calculate 60 pairs of unidirectional IDTs with 500~nm air-gap can enhance the external coupling to $\Gamma_{\mathrm{ext}}/2\pi=0.90$~kHz.   

Based on these improvements, the expected system parameters are listed in Table~S1.  
\begin{table}[h]
	\caption{\label{tab:table1}%
		Expected system parameters for a LN sphere with $D=330$~$\mu$m.
	}
		\begin{tabular}{lcdr}
			\textrm{Parameter}&
			\textrm{Value}\\
			\colrule
			Optical parameters & \\
			\colrule
			$Q_{\mathrm{opt}}$ & $5\times10^7$\\
			$\omega_{\mathrm{opt}}/2\pi$ & 300~THz\\
			$\kappa_{\mathrm{int}}/2\pi$ & 6~MHz\\
		    $\gamma_{\mathrm{int}}/2\pi$ & 6~MHz\\
			\colrule
			SAW parameters & \\
			\colrule
			$Q_{\mathrm{SAW}}$ & $5\times10^4$\\
			$\Omega/2\pi$ & 274~MHz\\
			$\Gamma_{\mathrm{int}}/2\pi$ & 5.5~kHz\\
			$\Gamma_{\mathrm{ext}}/2\pi$ & 0.90~kHz\\
			$\Gamma_{\mathrm{ext}}/\Gamma_{\mathrm{int}}$ & 0.16\\
			\colrule
			Photoelastic coupling & \\
			\colrule
			$g_0/2\pi$ & 2.00$\times 10^2$ Hz\\
			\colrule
		\end{tabular}
\end{table}
Using these parameters, we calculate the conversion performance with respect to the piezoelectric  external coupling for an optical input power of $P=3~\mu$W.  For the calculation, we assume a critical coupling ($\gamma_{\mathrm{in}}=\gamma_{\mathrm{ex}}$), which set the maximum obtainable conversion efficiency $\eta=0.5$. The system also satisfy the triple-resonance condition. $C_0$ and $\eta_{\mathrm{int}}=\frac{4C}{(1+C)^2}$ are the single photon cooperativity and the internal conversion efficiency, respectively.   We estimate a maximum conversion efficiency of $\eta=6.9\times10^{-2}$ at 	$\Gamma_{\mathrm{ext}}/\Gamma_{\mathrm{int}}=0.16$.

The numerical results are to be compared with the recent activities compiled by W.~Jiang~\textit{et al.}~\cite{Jiang2019}.   The relatively large device size compared to other studies comes in price with a small $C_0$. However, it is largely compensated by the high Q-factor, resulting in relatively large cooperativity and conversion efficiency with a small input power.  As a quantum transducer, it would be greatly beneficial if the SAW frequency can be in a GHz range, exceeding the typical frequency scale ($\sim 200$~MHz) of the thermal energy in the dilution refrigerator environment ($T=10$~mK).  The SAW frequency in this device is determined by the birefringence of the material, which only slightly changes with the geometrical alteration. 


\begin{thebibliography}{42}
\bibitem [{Sup()}]{Supplementary}%
	\BibitemOpen
	\href@noop {} {\bibinfo {title} {Appendix, which includes Refs.~[31-42] and further experimental
			and theoretical details.}}\BibitemShut {Stop}%
\end{thebibliography}
%
\end{document}